# Building the Chessboard-like Supramolecular Structure on Au (111) Surfaces

*Ruifen Dou[†] Yu Yang[‡] Ping Zhang[‡] Dingyong Zhong[§] Harald Fuchs[§] Yue Wang[∥] Lifeng Chi*[§]*

[†]Department of Physics, Beijing Normal University, Beijing 100875, China

[‡]LCP, Institute of Applied Physics and Computational Mathematics, Beijing 100088, China

[§]Physikalishes Institut, Westfälische Wihlms-Universität Münster, Wilhelm-klemm-Str. 10, 48149 Münster, Germany

[∥]Key Laboratory for Supramolecular Structure and Materials of the Ministry of Education, College of Chemistry, Jilin University, Chang Chun 130012, China

**ABSTRACT**: We investigate an anthracene derivative, 3(5)-(9-anthryl) pyrazole (ANP), self-assembled on the Au (111) surface by means of scanning tunneling microscopy (STM) and density functional theory (DFT) calculations. A chessboard-like network structure composed of ANP molecules is found, covering the whole Au (111) substrate. Our STM results and DFT calculations reveal that the formation of chessboard-like networks originates from a basic unit cell, a tetramer structure, which is formed by four ANP molecules connected through C-H…N hydrogen-bonds. The hydrogen bonds inside each tetramer and the molecular adsorption interaction are fundamentally important in providing a driving force for formation of the supramolecular networks.

The self-assembly of functional molecular building blocks into ordered supramolecular architectures on solid surfaces is believed as an effective route towards realization of possible application such as surface chemical functionalization, bio-sensors and molecular electronics.[1-5] Progress has been made in fabricating the ordered two-dimensional (2D) supramolecular structures on supported surfaces with desired patterns by using different molecules including small molecules with the symmetry configuration,[6-10] oligomers,[11-13] metal-supramolecular compounds,[14] molecular linkers coordinated with metals[15-18]. The self-assembled supramolecular architectures are predominately governed by a subtle balance between the intermolecular interactions and the molecule-substrate interaction. Generally, the intermolecule interaction including hydrogen bonds, metal coordination, dipole coupling, and van der Waals interactions are noncovalent, which result in the long-range ordered supramolecular aggregates. The molecule-substrate interaction determines the preferential adsorption sites on surface as well as molecular orientation related to the substrate lattice to form the commensurate supramolecular structures[19-21]. For those molecules which the strong and noncovalent intermolecular interactions such as hydrogen bonding, dipole-coupling or metal coordination interactions are easily produced between, the intermolecular interactions and molecular mobility play an important role in molecular packing characteristics on surfaces.[6, 22-23] Therefore, by carefully designing the molecular building blocks and choosing the suitable template surfaces to rationally adjust the intermolecule interactions and molecule-substrate, diverse supramolecular structures with considerable complexity, such as size-controlled porous networks[13,21,24-26], hierarchical structures[23, 27-29], chiral systems[29-35] and metal-coordination structures [15-18,36] have been achieved practically.

A recent report shows that organic molecule 3(5)-(9-anthryl) pyrazole (ANP) is designed as building blocks to construct different luminescent single crystals based on various combination of intermolecular hydrogen bonding and π-π stacking interaction.[37] Therefore, the optical properties of the obtained crystals can be tuned from blue, green, shallow blue up to blue-green due to the various intermolecular interactions mainly including π-stacking and hydrogen bonding. Thus it is promising to achieve the desirable ordered patterns upon deposition of ANP molecules on supporting surfaces through rationally adjusting the abundant interactions between the molecules. In addition, to further establish the relationship between different luminescent properties and the prepared supramolecular structures will enable the development of the high-performance organic electroluminescent devices. In the present paper, we report on the deposition of ANP on a Au (111) surface and the formation of a chessboard-like supramolecular structure driven by the combination of hydrogen-bonds in a tetramer and molecule-substrate interaction. The formation energy and the molecular packing model of a chessboard-like supramolecular structure are calculated according to the density functional theory (DFT).

The experiments were carried out in a UHV chamber equipped with a homemade molecular beam epitaxy (OMBE) system. The Au (111) crystal was cleaned by repeated cycles of Ar ion sputtering and annealing under ultrahigh vacuum (UHV) at a base pressure lower than $1.0 \times 10^{-10}$ Torr. Before the growth of the ANP monolayers, the organic materials was degassed and purified in vacuum by heating to its sublimation temperature (380 K) for several hours. ANP molecules were deposited onto a clean Au (111) surface by OMBE approach under UHV conditions. The Au (111) crystal was maintained at room temperature during deposition. All of the STM images were acquired at room temperature. In order to understand the electronic and geometrical details of the observed chessboard-like supramolecular structure, we performed first-principles calculations of the chessboard-like structure based on DFT. Computational details are provided in the Supporting Information.

Figure 1a shows molecular orbital (MO's) structures of an ANP calculated by DFT, which indicates that the plane angle between the pyrazole ring and the anthracene ring is 89.7˚. The spatial distribution of the highest occupied (HOMO) and lowest unoccupied molecular orbital (LUMO) of the ANP molecule are depicted in Figures 1b and 1c respectively. We can see here that both HOMO and LUMO mainly contribute around the C atoms of the anthracene ring, showing π orbital characters. Detailed wave-function analysis further reveals that they are composed of pz electrons of these C atoms. Figure 1d shows a clean Au (111) surface with reconstructed herringbone structures consisting of hexagonal close-packed (hcp) and face-centered cubic (fcc) regions and herringbone ridges and elbows.[38] Once monolayer ANP molecules are evaporated on the Au (111) surface, the molecules formed homogeneously ordered networks, as shown in Figure 1e. The reconstruction of the Au (111) surface can be clearly seen in Figure 1e. In addition, it is observed that one hole is constructed by four bright lobes, and the holes are arranged in a hexagonal way. In agreement with the threefold symmetry of the Au (111), three rotational domains of such structure coexist, which usually extend in the μm-range on the surface.

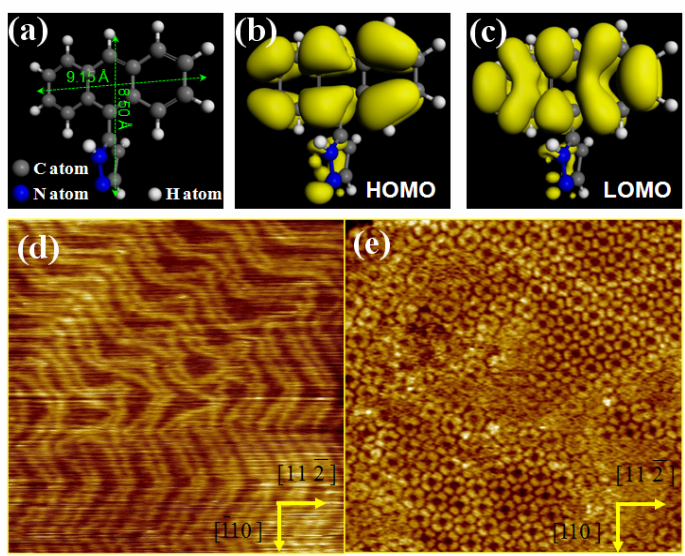

**Figure 1** (a) Schematic structure of 3(5)-(9-anthryl) pyrazole (ANP). The parallel and vertical length of an ANP molecule is 9.15 and 8.50 Å. (b) and (c) Highest occupied (HOMO) and lowest unoccupied molecular orbitals (LUMO) calculated by the density of functional theory (DFT). (d) STM image of a clean Au(111) surface with the herringbone reconstruction. (Size = 100 ×100 nm$^2$, V = 0.9 V, I = 5 nA ) (b) Typical large-scale STM image of ANP molecules adsorbed on Au(111) surfaces (Size = 25 ×25 nm$^2$, V = - 0.7 V, I = 0.1 nA ).

A typical zoom-in image is given in Figure 2a, which is obtained at the sample voltage of - 0.7 V. The STM image directly displays the real-space configuration of individual molecules and the periodic structure of the organic monolayers. It is obviously seen that an anthracene ring in an ANP molecule is resolved as a bright ellipse, which is very similar to the spatial distribution of the HOMO orbital shown in Figure 1b. Intuitively, four molecules make up of a tetramer, as depicted by gray molecular structures overlapped in Figure 2a. We deduce that the driving force for formation of a tetramer structure is the C-H…N hydrogen bonds between four pyrazole rings, which construct a hole. Owing to the low electron density in the pyrazole ring shown by the LOMO and HOMO of the ANP molecule (Figure 1b and 1c), four pyrazole groups connected by four hydrogen- bond display dark in the STM images. Here in order to form the hydrogen bonds between two neighboring pyrazole groups, the ANP molecules change from a nonplanar molecule into a planar one upon adsorption on the Au (111) surface. Moreover, the ANP molecules lying flat in the substrate surface might enhance the molecule-substrate interaction due to the π-orbital of ANP molecule coupling with the metal substrate. The length of C-H…N hydrogen bonds is approximately 2.24 Å, which is well corresponding to the previous report about hydrogen bonds.[39] The structure model of a tetramer structure is given in Figure 2c, in which the hydrogen bonds are denoted by the red lines. The tetramers regularly arrange themselves into the chessboard-like network structure on the whole Au (111) surface.

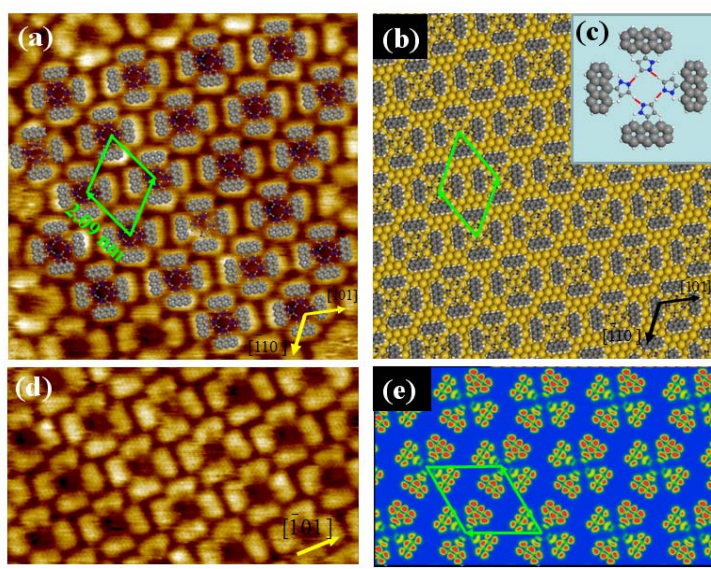

**Figure 2** (a) High-resolution STM image of ANP molecules self-assembled on Au(111) (Size = 16 ×16 nm$^2$, V = - 0.7 V, I = 0.1 nA ).The unit cell with a lattice constant of 2.09 ± 0.05 nm is indicated by the green rhombus. The ANP molecules corresponding to the bright bars are depicted. (b) The structure model of ANP self-assembled on Au (111) simulated by DFT. (c) A structure model of a tetramer structure formed through four hydrogen bonds (labeled by the red line). (d) and (e) The experimentally observed and theoretically simulated STM image of the adsorption structure of tetramers on the Au(111) surface. The experimental bias voltage is 0.7 V and the area size is 12.5 ×6.5 nm$^2$. The theoretical image shows the electronic state distribution integrated from -0.7 eV below to the Fermi energy, in a parallel plane 2.85 Å above the tetramer structure layer.

The orientation of the Au (111) surface is denoted by the two yellow arrows in Figure 2a. In the periodic chessboard-like networks, the center of symmetry of a tetramer is located along a [$\bar{1}10$] direction of the Au lattice. In the light of the STM results, we carefully ascertain the unit cell of the networks. It tells that two base vectors, $\vec{b}_1$ and $\vec{b}_2$, of the ordered suprastructure are the same and about 2.09 nm. The transfer matrix of the supramolecular structures can be calculated according to the lattice vectors of the unit cell of Au (111) and the structure and the orientation relationship between adsorbed molecules and the substrate lattice. Hence, the matrix of suprastructure relative to the Au (111) surface structure is as follows:

$$\begin{pmatrix} \vec{b}_1 \\ \vec{b}_2 \end{pmatrix} = \begin{pmatrix} 8.0 & 4.0 \\ 4.0 & 8.0 \end{pmatrix} \begin{pmatrix} \vec{a}_1 \\ \vec{a}_2 \end{pmatrix}$$

where $\vec{b}_{1,2}$ and $\vec{a}_{1,2}$ are the base vectors of the porous suprastructure and the Au (111) surface ($a_1 = a_2 = 0.289$ nm), respectively. The angle between $\vec{b}_1$ and $\vec{b}_2$ is 60°. Each unit cell contains four molecules and the area per molecule is 9.91 nm$^2$, which is reasonable compared with the molecular area of an individual planar ANP molecule. The matrix of supramolecular structure is integer, which implies formation of the commensurate structure. The schematic model of the chessboard-like network adsorbed on the Au (111) surface is shown in Figure 2b. Conspicuously, ANP molecules as planar molecules flat lie on the Au (111) surface to form a tetramer structure connected by four hydrogen bonds. The symmetry centers of tetramers are oriented along a [$\bar{1}10$] direction of the Au lattice. This means that the molecule-substrate interaction can not be negligible for the formation of the commensurate supramolecular structure [19-21].

To further validate our structural model of the chessboard-like network structure, we then perform simulations for the tetramer adsorption structure. The simulations are carried out following Tersoff and Hamann[40] and performed within density functional theory using the Vienna ab initio simulation package (VASP).[41] The first-principle calculation details are enclosed in DFT calculations section (see Supporting information). In this approach, the tunneling current is proportional to surface density of state summed over between surface Fermi energy and tip voltage. The typical STM result for the same area in Figure 2a at 0.7 V is shown in Figure 2d. Comparatively the corresponding results calculated at the tip voltage of -0.7 and 0.7 V are nearly the same and shown in Figure 2e. We can see that the theoretical results accord well with our experiments. The bright part of the tetramer structure corresponds to the π orbitals formed by $p_z$ electrons of the C atoms in the anthracene ring.

In order to further understand the driving force for formation of the chessboard-like porous networks, we examine the adsorption energy of an ANP molecule and the tetramer structure on the Au (111) surface using DFT calculations. Geometry optimizations find that an adsorbed ANP molecule prefers to lie flat on the Au (111) surface without any direct electronic overlapping. Figures 3a and 3c show the adsorption structure of an ANP molecule in its ground state and in the activated planar state respectively. The perpendicular distance between the anthracene ring and the Au (111) surface is found to be nearly the same for these two adsorption structures.

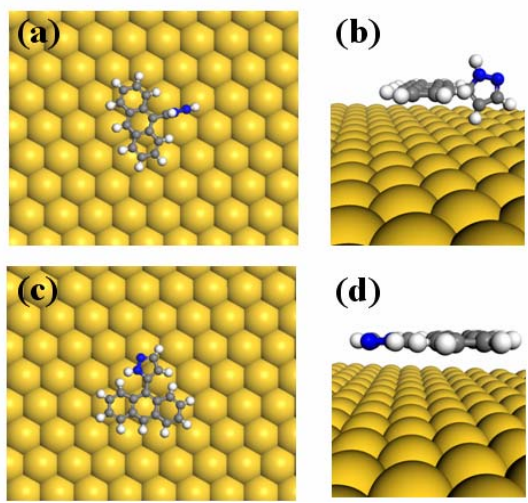

**Figure 3** (a) and (b) Top and side view of the adsorption structure of an ANP molecule in its ground state on the Au(111) surface. (c) and (d) Top and side view of the adsorption structure of a planar ANP molecule on the Au(111) surface.

However, changing the pyrazole ring from perpendicular to parallel to the Au surface enlarges the adsorption energy, which induces the molecular dipole moment. The calculated adsorption energy for the planar molecule is 0.15 eV, larger than the 0.09 eV adsorption energy of an ANP molecule in its ground state. Since the energy of an isolated ANP molecule in the planar state is 0.73 eV higher than that in its ground state, the 0.06 eV difference in adsorption energy is not large enough to activate an ANP molecule into the planar state. However, neighboring ANP molecules in the planar state can form C-H…N hydrogen bonds to further lower down the free energy. Four C-H…N hydrogen bonds can perfectly connect four ANP molecules together to form a tetramer structure, as shown in Figure 2c. The energy reduction per hydrogen bond is calculated to be 0.64 eV, which is close to the energy needed to activate an ANP molecule into the planar state. We further calculate the adsorption of tetramer structures on the Au (111) surface, and find that the adsorption energy per tetramer unit is 0.81 eV, which is larger than the adsorption energy of four isolated ANP molecules. The adsorption energy results indicate that the planar adsorption induced the dipole moment is energetically larger for tetramers than that for isolated molecules. More importantly, the adsorption energy enhancement per tetramer unit (0.81-0.09*4 = 0.45 eV) plus the energy reduction of four C-H…N hydrogen bonds (4×0.64 = 2.56 eV) is larger than the activation energy needed for transferring four ANP molecules into the planar state. Therefore, upon adsorption of the ANP molecules on the Au (111) surface, formation of the robust tetramer structures is energetically favorable.

In summary, we have investigated a light-emitting, nonplanar molecule of ANP self-assembled on Au (111) surfaces and formed ordered chessboard-like porous networks by STM observation and DFT calculations. Tetramer structures, created by four C-H…N hydrogen bonds between four pyrazole bases for each one, are found in the homogenous molecular film on

Au (111) surface. The long-range ordered chessboard-like porous networks consisted by ANP tetramers are driven by the strong substrate-molecule interaction including the dipole moment. The results of the present study fundamentally clarify the formation mechanisms of the supramolecular structures down to the single molecular level, which will perhaps rebound to understand how the molecular packing structures determine the electronic and optical properties. This will be furthermore helpful to design new high-performance organic electroluminescent devices.

**Supporting Information.**

This material is available free of charge via the Internet at http://arxiv.org.

**Corresponding Author**

\* chi@uni-muenster.de

ACKNOWLEDGMENT

The authors gratefully acknowledge support by the National Nature Science Foundation of China (Grant Nos. 10804010, 10904004, 0921003).